\documentstyle[aps,epsfig,floats,amssymb]{revtex}
\draft

\begin{document}
\twocolumn[\hsize\textwidth\columnwidth\hsize\csname
@twocolumnfalse\endcsname

\title{Chiral freedom and the scale of weak interactions}

\author{C. Wetterich}

\address{
Institut f{\"u}r Theoretische Physik,
Philosophenweg 16, 69120 Heidelberg, Germany}

\maketitle

\begin{abstract}
A second rank antisymmetric tensor field is proposed as an alternative to the Higgs scalar. No mass term is allowed by the symmetries. At the scale where the asymptotically free chiral couplings of the fermions grow large, condensates of top-antitop and bottom-antibottom may induce the spontaneous breaking of the electroweak gauge symmetry. This would solve the gauge hierarchy problem by dimensional transmutation, similar to QCD. 
\end{abstract}
\pacs{}

 ]

Convincing experimental evidence has established that the electroweak interactions can be described by a ``spontaneously broken gauge symmetry'' \cite{GSW}. The ``order parameter'' is a Lorentz-scalar and transforms as a doublet of weak isospin. The symmetries imply that the masses of the quarks and charged leptons are all proportional to the electroweak order parameter $\langle\varphi\rangle\approx~174~GeV$, just as the masses of the $W$- and $Z$-bosons. The standard ``Higgs mechanism'' \cite{H} assumes that the order parameter arises from the expectation value of a ``fundamental'' scalar field $\varphi(x)$, the Higgs scalar. The fermion masses - which contain the ``chiral information'' of the theory - are determined by the Yukawa couplings of the Higgs scalar. In turn, the Fermi scale $\langle\varphi\rangle$ is set by a free parameter of the theory, i. e. the mass term $\mu^2_\phi$ in the effective scalar potential. Its small size is difficult to understand within a unified theory.

The principal idea of this note relies on the observation that the chiral information can also be carried by an antisymmetric tensor field $\beta_{mn}(x)$. The renormalizable interactions between the fermions $\psi$ and the gauge bosons preserve a large global flavor symmetry. The observed fermion masses imply that this flavor symmetry must be broken by couplings connecting the left- and right-handed fermions $\psi_L$ and $\psi_R$ which belong to the representations $(2,1)$ and $(1,2)$ of the Lorentz symmetry. Bilinears involving $\psi_L$ and $\bar{\psi}_R$ transform as $(2,1)\otimes(2,1)=(3,1)+(1,1)$ and similar for $\psi_R\otimes\bar{\psi}_L=(1,3)+(1,1)$. The violation of the flavor symmetry may therefore either involve the Higgs scalar $(1,1)$ or the antisymmetric tensor $\beta_{mn}~(3,1)+(1,3)$. In fact, the six components of $\beta_{mn}=-\beta_{nm}$ can be decomposed into two inequivalent three dimensional irreducible representations of the Lorentz group
\begin{equation}\label{1}
\beta^\pm_{mn}=\frac{1}{2}\beta_{mn}\pm\frac{i}{4}\epsilon_{mn}\ ^{pq}\beta_{pq},
\end{equation}
with $\epsilon_{mnpq}$ the totally antisymmetric tensor $(\epsilon_{0123}=1)$. We observe that the representations $(3,1)$ and $(1,3)$ are complex conjugate to each other. (The Lorentz-indices $m,n...$ are raised by $\eta^{mn}=diag(-1,+1,+1,+1)$.) 

The complex fields $\beta^\pm_{mn}$ should belong to doublets of weak isospin and carry hypercharge $Y=1$. The explicit breaking of the flavor symmetry can then be encoded in the ``chiral couplings'' $\bar{F}_{U,D,L}$, i.e.
\begin{eqnarray}\label{2}
-{\cal L}_{ch}&=&\bar{u}_R\bar{F}_U\tilde{\beta}_+q_L-\bar{q}_L
\bar{F}^\dagger_U\stackrel{\eqsim}{\beta}_+u_R\nonumber\\
&&+\bar{d}_R\bar{F}_D\bar{\beta}_-q_L-\bar{q}_L\bar{F}^\dagger_D\beta_-d_R\nonumber\\
&&+\bar{e}_R\bar{F}_L\bar{\beta}_-l_L-\bar{l}_L\bar{F}^\dagger_L\beta_-e_R.
\end{eqnarray}
Here $u_R$ is a vector with three flavor components corresponding to the right handed top, charm and up quarks and similar for $d_R$ and $e_R$ for the right handed $(b,s,d)$ and $(\tau,\mu, e)$ or $q_L,l_L$ for the three flavors of left handed quark and lepton doublets. Correspondingly, the chiral couplings $\bar{F}_{U,D,L}$ are $3\times 3$ matrices in generation space. We have used the short hands $(\sigma^{mn}\gamma^5 =\frac{i}{2}\epsilon^{mn}\ _{pq}\sigma^{pq})$ 
\begin{eqnarray}\label{3}
\beta_\pm &=&\frac{1}{2}\beta^\pm_{mn}\sigma^{mn}=\frac{1}{2}\beta_{mn}\sigma^{mn}_\pm 
=\beta_{\pm}\frac{1\pm\gamma^5}{2}\nonumber\\
\bar{\beta}_\pm&=&\frac{1}{2}(\beta^\pm_{mn})^*\sigma^{mn}
=D^{-1}\beta^\dagger_\pm D =\bar{\beta}_\pm \frac{1\mp\gamma^5}{2}\nonumber\\
\tilde{\beta}_+&=&-i\beta^T_+\tau_2~,~ \stackrel{\eqsim}{\beta}_+=i\tau_2\bar{\beta}_+
\end{eqnarray}
with $\sigma_{\pm}^{mn}=\frac{1}{2}(1\pm\gamma^5)\sigma^{mn},~\sigma^{mn}=\frac{i}{2}
[\gamma^m,\gamma^n]$ and $\bar{\psi}=\psi^\dagger D,~D=\gamma^0,~\gamma^5=-i\gamma^0\gamma^1\gamma^2\gamma^3,~\psi_L=\frac{1}{2}(1+\gamma^5)\psi$. The transposition $\beta^T$ and $\tau_2$ act in weak isospin space, i.e. on the two components of the weak doublet $\beta^+_{mn}$.

The Lagrangian (\ref{2}) accounts for the most general renormalizable interactions between $\beta$ and the fermions which are consistent with the gauge symmetries. Beyond conserved baryon number $B$ and lepton number $L$ it is further invariant with respect to a global axial $U(1)_A$ symmetry with charges $A=1$ for the right handed fermions, $A=-1$ for the left handed ones, $A=-2$ for $\beta^-$ and $A=2$ for $\beta^+$. Such an axial symmetry has an important consequence: no mass term is allowed for the fields $\beta^\pm$! Indeed, a Lorentz singlet is contained in $\beta^+\beta^+$ or $\beta^+(\beta^-)^*$, the first being forbidden by hypercharge and the second by axial symmetry. More generally, mass terms are forbidden by any symmetry under which $\beta^+$ and $\beta^-$ transform differently such that $\beta^+(\beta^-)^*$ is not invariant. An example is a discrete $Z_N$-subgroup of $U(1)_A, N>4$. The minimal symmetry $G_A$ is an axial $Z_2$ symmetry where $\beta_-$ and $d_R,e_R$ are odd whereas all other fields have even parity. We will assume that all interactions are invariant under the  $G_A$-symmetry.

The absence of an allowed mass term is in sharp contrast to the Higgs mechanism and constitutes  a crucial part of the solution of the hierarchy problem. On the other hand, there exist kinetic terms $\sim(\beta^+)^*\beta^+$ and $(\beta^-)^*\beta^-$. The most general one allowed by all symmetries reads 
\begin{eqnarray}\label{4}
-{\cal L}^{ch}_{kin}&=&-(D_\mu\beta^{+\mu\rho})^\dagger D_\nu\beta^{+\nu}{_\rho}-(D_\mu\beta^{-\mu\rho})^\dagger D_\nu\beta^{-\nu}{_\rho}.
\end{eqnarray}
Here $D_\mu$ is the covariant derivative involving the gauge fields of $SU(2)_L\times U(1)_Y$ and the gravitational spin connection and $e^m_\mu$ denotes the vierbein. The action $S_M=\int d^4xe{\cal L}, ~e=\det(e^m_\mu)=\sqrt{g}$, involves in addition the usual covariant kinetic term for the fermions and gauge fields of the standard model, as well as gravitational parts. A kinetic term of this type has been discussed in the context of conformal supergravity \cite{WH} (see also \cite{CH}). It does not admit an additional local gauge symmetry. 

Finally, dimensionless quartic couplings appear in the form 
$(\beta^\dagger\beta)^2, (\beta^\dagger\vec{\tau}\beta)^2$ with appropriate contractions of Lorentz indices and an even number of factors involving $\beta^+$ or $\beta^-$. Couplings of the type 
$\big((\beta^+)^\dagger\beta^-\big)\big((\beta^+)^\dagger\beta^-\big)$ violate the continuous $U(1)_A$ symmetry, while the  discrete $G_A$ symmetry $\beta^-\rightarrow -\beta^-$ is preserved \cite{CWH}.

The sign of the kinetic term is fixed by the positivity of the energy density
\begin{eqnarray}\label{4a}
\rho=-T^0_0=&&\{\partial_0B^{k*}_+\partial_0B^k_++2\partial_kB^{k*}_+\partial_lB^l_+\nonumber\\
&&-\partial_lB^{k*}_+\partial_lB^k_+\}+(+\rightarrow -)
\end{eqnarray}
where
\begin{equation}\label{4b}
B^k_\pm=\pm i\beta^{0k}_\pm=\frac{1}{2}\epsilon^{kmn}\beta_{\pm mn}.
\end{equation}
(The energy momentum tensor $T^{\mu\nu}$ obtains from variation of $S_M$ with respect to the vielbein.) A plane wave of $\beta_+$ (or $\beta_-$) with momentum $\vec{q}=(0,0,q)$ may be decomposed into transversal and longitudinal modes 
$\vec{B}=\big(\frac{1}{\sqrt{2}}(b_1+b_2),-\frac{i}{\sqrt{2}}(b_1-b_2),b_3\big)$ which obey the field equations
\begin{equation}\label{4c}
(q_0+q)^2b_1=0,~(q_0-q)^2b_2=0,~(q^2_0-q^2)b_3=0.
\end{equation}
Only the longitudinal component contributes to $\rho$
\begin{equation}\label{4e}
\rho=2\partial_kb^*_3\partial_kb_3.
\end{equation}

With respect to plane waves the free theory is on the borderline between stability and instability - the mass and the kinetic energy for $b_{1,2}$ both vanish. Beyond the plane wave solutions the free theory admits secular solutions that reflect the unboundedness of the free Hamiltonian \cite{CWCS}. In a borderline situation for the free theory the interactions will decide on which side of the divide the interacting theory lies. We argue in \cite{CWH} (see also below) that the interactions generate a non-perturbative mass term that renders our model stable. All solutions of the field equations derived from the quantum effective action have then positive energy density. The particles corresponding to the chiral antisymmetric tensor fields are stable massive spin one particles, or resonances if their decay into lighter particles is included.

The chiral couplings are asymptotically free - in contrast to the Yukawa couplings of the Higgs scalar. We therefore call our scenario ``chiral freedom''. The remarkable feature is the opposite sign of the fermion anomalous dimensions for the contribution from $\beta_\pm$ as compared to the one from the Higgs scalar. We introduce an infrared cutoff scale $k$ \cite{CWERGE} (or equivalently a renormalization scale $\mu$ defined by nonvanishing ``external momenta''). Neglecting effects involving the electroweak gauge couplings the running of the renormalized chiral couplings obeys
\begin{eqnarray}\label{4f}
k\frac{\partial}{\partial k}F_U&=&-\frac{9}{8\pi^2}F_UF^\dagger_UF_U-
\frac{3}{8\pi^2}F_UF^\dagger_DF_D\nonumber\\
&&+\frac{1}{4\pi^2}F_U ~tr(F^\dagger_UF_U)-\frac{1}{2\pi}^2g^2_sF_U\nonumber\\
k\frac{\partial}{\partial k}F_D&=&-\frac{9}{8\pi^2}F_DF^\dagger_DF_D-
\frac{3}{8\pi^2}F_DF^\dagger_UF_U\\
&&+\frac{1}{4\pi^2}F_D tr(F^\dagger_DF_D+\frac{1}{3}F^\dagger_LF_L)-
\frac{1}{2\pi^2}g^2_sF_D\nonumber\\
k\frac{\partial}{\partial k}F_L&=&-\frac{9}{8\pi^2}F_LF^\dagger_LF_L+
\frac{1}{4\pi^2}F_L ~tr(F^\dagger_DF_D+\frac{1}{3}F^\dagger_LF_L)\nonumber
\end{eqnarray}
There are no vertex corrections from $\beta$-exchange diagrams. 

Very similar to QCD, chiral freedom comes in pair with a characteristic infrared scale $\Lambda_{ch}$ where the chiral couplings grow large. Let us concentrate on the single chiral coupling of the top quark $f_t$ and neglect all other chiral couplings and $g_s$ for the moment. The solution of the evolution equation (\ref{4f}) 
\begin{equation}\label{8}
f^2_t(k)=\frac{7\pi^2}{4\ln(k/\Lambda^{(t)}_{ch})}
\end{equation}
exhibits by dimensional transmutation the ``chiral scale'' $\Lambda^{(t)}_{ch}$ where $f_t$ formally diverges. We propose that $\Lambda^{(t)}_{ch}$ sets also the characteristic mass scale for electroweak symmetry breaking. The solution of the gauge hierarchy problem is then very similar to the understanding of a small ``confinement scale'' $\Lambda_{QCD}$ in a setting with fundamental scales given by some grand unified scale or the Planck scale.

Indeed, we suggest that the strong coupling $f_t$ induces a top-antitop condensate $\langle\bar{t}_Lt_R\rangle\neq 0$. Here $\bar{t}_Lt_R$ plays the role of a composite Higgs field \cite{BL} - a nonvanishing expectation value leads to ``spontaneous breaking'' of the electroweak gauge symmetry, giving the $W$- and $Z$-bosons a mass while leaving the photon massless. The composite Higgs doublet fields $\varphi_t\sim\bar{q}_Lt_R,~\varphi_b\sim\bar{b}_Rq_L$ carry the same electroweak quantum numbers as the usual ``fundamental'' Higgs field. If $\varphi_t$ and $\varphi_b$ are normalized with a standard (gauge invariant) kinetic term the observed masses of the $W,Z$-bosons follow for a Fermi scale $\langle\varphi\rangle=(\langle\varphi_t\rangle^*\langle\varphi_t\rangle +\langle\varphi_b^*\rangle \langle\varphi_b\rangle)^{1/2}=174~GeV$. With $\langle\varphi\rangle=a\Lambda_{ch}$ (where $a$ remains to be computed) this can be achieved by a suitable value of $f_t$ at some appropriate short distance scale. The symmetries allow for effective Yukawa couplings of $\varphi_t$ to the up-type quarks and $\varphi_b$ to the down-type quarks and charged leptons. With respect to the axial 
$G_A$-symmetry $\varphi_t$ and $\varphi_b$ carry the same charges as $\beta_+$ and $\beta_-$. For $\langle\varphi_t\rangle\neq 0,~\langle\varphi_b\rangle\neq 0$ all quarks and charged leptons can acquire a mass.

For a first check if these ideas are reasonable we solve the coupled gap equations for the top-quark- and $\beta^+$-propagators. The lowest order Schwinger-Dyson equation \cite{SD} for a possible top quark mass $m_t$ reads 
\begin{equation}\label{9}
m_t=-4im^*_t\int\frac{d^4q}{(2\pi)^4}
\frac{R_{mn}(q)\delta^{mn}f^2_t(q)}{q^2+|m_t|^2}
\end{equation}
with $R_{mn}(q)$ the $\beta^+-\beta^+$component of the $\beta$-propagator-matrix. For $m_t=0$ one would have $R_{mn}(q)=0$ by virtue of conserved hypercharge. However, $U(I)_Y$ is spontaneously broken by $\langle\varphi_t\rangle\neq 0$ or $m_t\neq 0$. By inserting a top-quark loop we obtain the lowest order Schwinger-Dyson equation for the inverse propagator of the electrically neutral component of $\beta^+$ and infer
\begin{eqnarray}\label{10}
&&R_{mn}(q)\delta^{mn}=\frac{72i\big(f^*_t(q)\big)^2 I(q)}{q^4},\\
&&I(q)=\int \frac{d^4q'}{(2\pi)^4}
\frac{m^2_t}{[(q'+\frac{q}{2})^2+|m_t|^2][(q'-\frac{q}{2})^2+|m_t|^2]}.\nonumber
\end{eqnarray}
Alternatively, we could integrate out the fields $\beta^\pm$ in favor of a non-local four-fermion interaction. In this purely fermionic formulation (besides the gauge interactions) we have recovered the term corresponding to the combination of eqs. (\ref{9})(\ref{10}) in the Schwinger-Dyson equation at two loop order. 

For eqs. (\ref{9})(\ref{10}) we have made several simplifications. (i) We have replaced an effective momentum dependent mass term $m_t(p)$ by a constant even though $m_t(p)$ is supposed to vanish fast for large $p^2$. This momentum dependence would provide for an effective ultraviolet cutoff in the $q'$-integral for $I(q)$. We take this effect into account by approximating (with $\Lambda_t$ a suitable UV-cutoff)
\begin{equation}\label{11}
I(q)=\frac{im^2_t}{32\pi^2}\ln
\frac{\Lambda^4_t+(|m_t|^2+q^2/4)^2}{(|m_t|^2+q^2/4)^2}.
\end{equation}
(ii) We have used the classical $\bar{\beta}^+-\beta^+$ propagator for the inversion of the two point function, as needed for the computation of $R_{mn}$. This results in an IR-divergence due to the squared propagator $\sim q^{-4}$. Presumably, the full propagator will contain a mass term acting as an IR-regulator (see below). We take this into account by limiting the $q$-integral in eq. (\ref{9}) to $q^2>M^2_\beta>\Lambda^2_{ch}$. Combining eqs. (\ref{9}-\ref{11}) results in the condition $(m_t\hat{=}|m_t|)$
\begin{equation}\label{12}
\int^\infty_{M^2_\beta/m^2_t}\frac{dx}{x(x+1)}
\frac{\ln\left(\frac{\Lambda^4_t}{m^4_t(1+x/4)^2}+1\right)}
{\left(\ln\left(\frac{m^2_t}{\Lambda^{(t)}_{ch^2}}\right)+\ln x\right)^2}
=\frac{49}{36}
\end{equation}
For not too large $M^2_\beta/m^2_t$ this has indeed a nontrivial solution with $m_t>\Lambda^{(t)}_{ch}$. For the example $\Lambda_t=3m_t,M_\beta=m_t$ one finds $m_t=1.45~ \Lambda^{(t)}_{ch}$.

We conclude that the generation of a nonvanishing top-antitop condensate $\langle\varphi_t\rangle\neq 0$ seems plausible. From a different viewpoint, the gap equations have a nontrivial solution if the chiral coupling $f_t$ exceeds a critical value. This is always the case since $f_t$ grows to very large values unless an infrared cutoff $\sim m_t$ is generated by electroweak symmetry breaking. Our result corresponds to $f_t(m_t)=6.8$. A similar gap equation for the $W$-boson mass could determine $M_W/\Lambda^{(t)}_{ch}$ - in consequence the ratio $m_t/M_W$ would become predictable! Its computation amounts to the determination of the effective top quark Yukawa coupling $h_t=m_t/\langle\varphi_t\rangle$. 

We will assume that both $\varphi_t$ and $\varphi_b$ acquire nonvanishing expectation values such that all particles acquire a mass proportional to the Fermi scale except the photon and the neutrinos. One possible mechanism for inducing $\langle\varphi_b\rangle\neq 0$ is the continued running of the chiral coupling $f_b$. Its increase is not stopped by $\langle\varphi_t\rangle\neq 0$ and may result in a second scale $\Lambda^{(b)}_{ch}$. Alternatively, effective interaction terms of the type $(\varphi^\dagger_t\varphi_b)(\varphi^\dagger_b\varphi_t)$ may trigger expectation values of a similar size $\langle\varphi_b\rangle\approx\langle\varphi_t\rangle$. The quark- and charged lepton  masses are then generated by effective Yukawa couplings $U,D,L$ of the composite scalar fields
\begin{eqnarray}\label{12a}
-{\cal L}_y=\bar{u}_RU\tilde{\varphi}_tq_L
&-&\bar{q}_LU^\dagger\tilde{\varphi}^\dagger_tu_R\nonumber\\
+\bar{d}_RD\varphi^\dagger_bq_L&-&\bar{q}_LD^\dagger\varphi_bd_R\nonumber\\
+\bar{e}_RL\varphi^\dagger_bl_L&-&\bar{l}_LL^\dagger\varphi_be_R
\end{eqnarray}
according to $M_U=U\langle\varphi_t\rangle, M_D=D\langle\varphi_b\rangle^*,M_L=L\langle\varphi_b\rangle^*$. 

Possible generation and mixing patterns in the chiral couplings $F_{U,D,L}$ will be reflected in the effective fermion masses. It is always possible to make $F_{U,D,L}$ diagonal and real by appropriate chiral transformations. This results in a Cabibbo-Kobayashi-Maskawa-type mixing matrix for the weak interactions and renders issues like particle decays, strangeness violating neutral currents or CP-violation very similar to the Higgs-mechanism in the standard model. Up to the small weak interaction effects the diagonal $F_{U,D,L}$ imply separately conserved quantum numbers for the different fermion species - in turn also the mass matrices $M_{U,D,L}$ must be diagonal in this basis. The transmission of generation structures from the chiral couplings $F$ to the mass matrices $M$ can be understood in terms of symmetries. As an example, consider diagonal $F_{U,D}$ with $f_t$ and $f_b$ as the only nonvanishing entries. Omitting again weak interactions this would result in an enhanced global chiral $SU(4)_L\times SU(4)_R$-symmetry acting on $u,d,s,c$. Such a flavor symmetry would not be affected by nonzero $\
\langle\bar{t}_Lt_R\rangle,\langle\bar{b}_Rb_L\rangle$ and forbids mass terms for the four ``light'' quark flavors. Switching on $f_c$ (and/or off diagonal $F_{tc},F_{ct}$) allows for the generation of a charm quark mass. Now the flavor symmetry is reduced to $SU(3)_L\times SU(3)_R$ and we recover the usual setting of chiral symmetries in QCD with three massless quarks, including the anomalous axial symmetry. Similar considerations hold for the different scales of $m_s$ and $m_{u,d}$ and for the hierarchies of the charged lepton masses.

The strong chiral interactions can also lead to the generation of mass terms for the chiral tensors. This issue is best understood in a field basis which reflects more closely the particle content for massive chiral tensor fields, namely massive spin one particles - the ``chirons''. With respect to the little group for massive particles $SO(3)$ the antisymmetric tensors are equivalent to vectors. We define
\begin{equation}\label{c4a}
S^{\pm\mu}=\frac{\partial_\nu}{\sqrt{\partial^2}}\beta^{\pm\nu\mu}~,~\partial_\mu S^{\pm\mu}=0
\end{equation}
such that the kinetic term (\ref{4}) reads
\begin{equation}\label{c4b}
-{\cal L}^{ch}_{kin}=(\partial^\nu S^{+\mu})^\dagger(\partial_\nu S^+_\mu)+(\partial^\nu S^{-\mu})^\dagger(\partial_\nu S^-_\mu).
\end{equation}
A non-perturbatively generated mass term 
\begin{equation}\label{c4c}
-{\cal L}^{ch}_M=m^2_+(S^{+\mu})^\dagger S^+_\mu+m^2_-(S^{-\mu})^\dagger S^-_\mu
\end{equation}
renders the free chirons classically stable. Indeed, the issue of stability is decided by the properties of the solutions of the field equations derived from the quantum effective action rather than the classical action. The presence of a mass term (\ref{c4c}) with positive $m^2_\pm$ implies a positive energy density for the asymptotic solutions. In this case our model has neither ghosts nor tachyons and the interactions have pushed the free theory of chiral tensors towards the side of stability. Including also an imaginary part in $m^2_\pm$ reflecting the decays into lighter particles, the chirons become massive spin one resonances. Their description is fully consistent, with a similar status in several respects as the massive $\rho$-mesons in QCD. We actually expect the width of the chirons to be rather broad. In view of the small couplings to the quarks of the first generation their detection at LHC will not be easy.

In terms of the original fields $\beta^\pm_{\mu\nu}$ the mass term (\ref{c4c}) is not local. It cannot be generated perturbatively. Nevertheless, nonvanishing mass terms proportional to the non-perturbative scale $\Lambda_{ch}$ (where the chiral couplings grow large) seem plausible \cite{CWH}. We also note that the spontaneous electroweak symmetry breaking induces new effective cubic couplings for the chiral tensors which would be forbidden by an unbroken gauge symmetry. In presence of such cubic couplings the loops with intermediate chiral tensors indeed generate a contribution to the mass term of the type (\ref{c4c}), with $\Delta m^2_\pm$ proportional to the square of the effective cubic coupling \cite{CWH}.

Let us close this note by a few remarks on the predictivity of our model and possible future tests. By parameter counting the entries in the effective quark and lepton mass matrices $M_{U,D,L}$ correspond to free parameters in the chiral couplings $F_{U,D,L}$. If no further relevant (or marginal) couplings beyond $f_t,g$ play a role, the ratio $M_W/m_t$ becomes a prediction. However, we note that it is possible that the renormalizable quartic couplings $\sim\beta^4$ (not discussed here) influence this relation. The effective masses of the composite scalar fields $\varphi_t,\varphi_b$ as well as the masses of $\beta^\pm$ are predictable. Radiative corrections and the detailed structure of effective vertices may be modified by the presence of the $\beta$-field. The strongest influence is expected in the $t$ and $b$-sectors due to the strong chiral couplings $f_t,f_b$. In \cite{CWH} we argue that the present observations impose a bound on the chiron mass of approximately $M_c\gtrsim300$ GeV. Of particular interest is the mixing between the chirons and the photons which results in a correction to the anomalous magnetic moment of the muon. A more detailed investigation of the effective interactions, for example by the use of functional renormalization \cite{CWERGE}, may reveal very interesting features of new physics. This should answer the question if chiral freedom results in an acceptable solution of the gauge hierarchy problem.

\bigskip
\noindent
{\bf Note added:} The author thanks M. V. Chizhov for sending his work on antisymmetric tensor fields \cite{CH} after the first e-print version of this letter and pointing out that asymptotic freedom has been found earlier in an abelian model for chiral tensors. Since the first version we have also extended the discussion of mass generation for chiral tensors, stability of our model as well as phenomenological implications \cite{CWH}, \cite{CWCS}.


\begin{thebibliography}{}
\bibitem{GSW}S. Glashow, Nucl. Phys. 22, 579 (1961);\\
A. Salam, in Elementary particle theory, ed. N. Svartholm, (Almquist and Wiksell, Stockholm, 1968);\\
S. Weinberg, Phys. Rev. Lett. 19, 1264 (1967)
\bibitem{H}P. W. Higgs, Phys. Lett. 12, 132 (1964)
\bibitem{WH}B. de Wit, J. W. van Holten, Nucl. Phys. {\bf B155}, 530 (1979); 
E. Bergshoeff; M. de Roo, B. de Wit, Nucl. Phys. {\bf B182}, 173 (1981)
\bibitem{CH}M. V. Chizhov, Mod. Phys. Lett. A8, 2753 (1993)\\
L. V. Avdeev, M. V. Chizhov, Phys. Lett. B321, 212 (1994)
\bibitem{CWH}C. Wetterich, hep-ph/0607051
\bibitem{CWCS}C. Wetterich, hep-th/0509210
\bibitem{CWERGE}C. Wetterich, Phys. Lett. B 301, 90 (1993); Z. Phys. {\bf C57}, 451 (1993)
\bibitem{BL}V. A. Miranski, M. Tanabashi, K. Yamawaki, Phys. Lett. B221, 177 (1989); Mod. Phys. Lett. A4, 1043 (1989)\\
W. Bardeen, C. Hill, M. Lindner, Phys. Rev. D41, 1647 (1990) 
\bibitem{SD}F. J. Dyson, Phys. Rev. 75, 1736, (1949);\\
J. S. Schwinger, Proc. Nat. Acad. Sci 37, 452 (1951)
\end{thebibliography}
\end{document}